\documentclass{elsart1p}

 \usepackage{graphics}
 \usepackage{graphicx}
 \usepackage{epsfig}

\usepackage{amssymb}

\usepackage{bbm}
\usepackage{epsfig}
\usepackage{amsmath}
\usepackage{amsfonts,amsbsy}
\usepackage{amssymb}
\usepackage[dvips]{color}

\def\q{{\boldsymbol q}}

\def\x{{\boldsymbol x}}

\def\balphat{\tilde {\boldsymbol \alpha}}

\def\alphat{{\tilde  \alpha}}

\def\betat{{\tilde  \beta}}

\def\At{{\tilde  A}}

\newcommand{\beq}{\begin{eqnarray}}

\newcommand{\eeq}{\end{eqnarray}}

\long\def\comment#1{ }    

\newcommand{\be}{\begin{equation}}

\newcommand{\ee}{\end{equation}}

\newcommand{\nn}{\nonumber\\ }

\newcommand{\labe}{\label}

\def\del{\partial}

\begin{document}

\begin{frontmatter}



\title{Classical initial conditions in high energy nucleus-nucleus collisions}


\author{Yacine Mehtar-Tani}

\address{Institut f\"ur Theoretische  Physik\\
Philosophenweg 16, D-69120 Heidelberg, Germany}

\begin{abstract}
An iterative proceedure is proposed to compute the classical gauge field produced in the collision of two heavy nuclei at high energy. The leading order is obtained by linearizing the Yang-Mills equations in the light-cone gauge, and provides a simple formula for gluon production in nucleus-nucleus collisions. At this order $k_t-$factorization breaks down.
\end{abstract}



\end{frontmatter}

\section{Introduction}
\label{intro}


Perturbative QCD, based on small coupling calculations, shows that the gluon density in the nuclear wave function rises quickly at small $x$ (large energy) and eventually saturates to preserve unitarity.  The saturation of the gluon distribution is governed by a single scale, the so-called saturation scale $Q_s\gg \Lambda_{QCD}$, which sets the hard scale for pertubative calculations. This idea is formalized in the recently developped theory of the color glass condensate (CGC)\cite{McLerV,JalilKLW,JalilKMW1,IancuLM,Balit1,Kovch3}. \\

In the framework of the CGC, the problem of gluon production reduces to solving the classical Yang-Mills equations with a classical current describing the fast moving sources. To calculate observables, one has to average over all possible configurations of the sources with a given statistical distribution. 
To probe the saturation regime, proton(deuteron)-nucleus collisions have been studied at RHIC since they represent a clean probe of the nuclear wave function at high energy. The gluon production cross-section has been calculated analytically \cite{KovchM3,KovnW,KovchT1,DumitM1,BlaizGV1,GM}.
In this case, the proton is a dilute object, whereas the nucleus is saturated. To do so, one linearizes the Yang-Mills equations with respect the weak proton field resuming all high density effects in the nucleus. $k_t-$factorization holds and allows one to express the cross-section as a convolution in transverse momentum space of proton and nucleus gluon distributions. Also, it has been shown that final state interactions are absent, therefore no medium is produced. 
    In nucleus-nucleus collisions, where both projectiles are in the saturation regime, numerical studies of gluon production in nucleus-nucleus collisions at high energy, based on the \cite{JKMW1,JKMW2} (see also \cite{GV}), have been performed \cite{KrasnV,Lappi1}, but no exact solutions have been found so far. Kovchegov was the first to consider analytically the problem, however the formula he proposes is derived by conjecturing the absence of final state interactions in addition to other assumptions \cite{KovAA}. Later, I. Balitsky calculated the first correction the proton-nucleus by expanding the gauge field symmetrically in powers of commutators of Wilson lines \cite{Balit2}.\\

 In this work \cite{BlaizM}, we propose an analytic method, in the CGC framework, to calculate the cross-section for gluon production in nucleus-nucleus collisions. 
\section{The classical gauge field in nucleus-nucleus collisions}
The Yang-Mills equations read
\be\label{Yang-Mills}
D_\mu F^{\mu\nu}=J^{\nu},
\ee
where $J^\nu$ is a conserved current describing the fast moving nuclei, A and B, moving respectively in the $+z$ and $-z$ direction. Their source densities $\rho_{_A}(x^+,\x)\sim \delta(x^+)\rho_{_A}(\x)$ and  $\rho_{_B}(x^-,\x)\sim\delta(x^-)\rho_{_B}(\x)$ are confined near the light-cone. $x^+=(t+z)/\sqrt{2}$ and $x^+=(t+z)/\sqrt{2}$ are the light cone variables, and $\x$ represents the transverse coordinate. As we shall see, the relevant degrees of freedom are  the Wilson lines :
\begin{eqnarray}
U(\x)&=&{\mathcal P} \exp\left(ig\int dz^+ \frac{1}{\del^2_\perp}\rho_{_A}(z^+,\x)\cdot T\right),\\
V(\x)&=&{\mathcal P} \exp\left(ig\int dz^- \frac{1}{\del^2_\perp}\rho_{_B}(z^-,\x)\cdot T \right),
\end{eqnarray}
for nucleus A, and nucleus B respectively.
 The case where one of the sources is weak ( corresponding to proton-nucleus collisions) has been solved analytically, however, the scattering of two dense systems is still an unsolved problem. In our work we present an iterative method to built the solution of the Yang-Mills equation for this problem with the aim to capture the main essence of the physics in the first iteration which, as we shall see, can bee easily computed. 
For this purpose, the light-cone gauge $A^+=0$ is very convenient, and leads to a lot of simplifications. In this gauge the current can be constructed exactly, and the field immediately after the collision follows easily from the yang-Mills equations. \\

The nuclei fields are not determined uniquely, because of the additional gauge degree of freedom in axial gauges. One simple configuration is $A^-_{_A}=-\frac{1}{\del^2_\perp}\rho_{_A}(x^+,\x)$, $A^i_{_A}=A^+_{_A}=0$, for nucleus A and  $A^{i}_{_B}=-\int dy^-V^\dag(y^-,\x)\;\frac{\del^i}{\del^2_\perp}\rho_{_B}(y^-,\x)$ for nucleus B. 
The field just after the collision is determined exactly for this initial condition $A^i=UA^i_{_B}$ and 
$\del^+A^-=\left(\del^iU\right) A_{_B}^i$.
The transverse pure gauge field of nucleus B is present all the way to $t=\infty$, this would lead to technical complication while computing the field in the forward light-cone. To avoid that one can simplify further by removing this pure gauge by a gauge rotation involving the gauge link $V$, i.e., $\At^\mu\cdot T=V^\dag (A^\mu\cdot T)V-\frac{1}{ig}V^\dag \del^\mu V$.
So the produced gauge field near the light-cone gets rotated leading to
\be
\At^i= V(U-1)A^i_{_B}\equiv\alphat_0^i, \;\;\;\del^+\At^-=V(\del^j U)A^j_{_B}\equiv\betat_0.\labe{beta0VU}
\ee
Having the exact field produced immediately after the collision of two heavy nuclei one can think of solving the equations of motion iteratively in powers of this initial field. 
Let us define an expansion in powers of the initial fields $\alphat_{_0}$ and $\betat_{_0}$:
\be
\At^\mu=\sum_{n=0}^{\infty}\At^\mu_{(n)}\;.\labe{expan}
\ee
In this case the zero order is simply obtained by gauge rotating the nuclei fields  $A^-_{_A}$ and $ A^{i}_{_B}$,
\be
\At^\mu_{(0)}=-\delta^{\mu-}V\frac{1}{\del^2_\perp}\rho_{_A}-\delta^{\mu+}\frac{1}{\del^2_\perp}\rho_{_B}\;,\labe{Ain}
\ee
Note that we have generated a $+$ component of the gauge field. Strictly speaking,  we switch to  $\At^+=-\frac{1}{\del^2_\perp}\rho_{_A}$ gauge, which reduces to $\At^+=0$ in the forward light-cone since the source has its support only on the light-cone, $x^+=0$. 
The equations of motion for the first order read
\beq
&&\del^+(\del_\mu\At^\mu_{(1)})=0,\labe{YMtlin+}\\
&& \square \At^-_{(1)}-\del^-\del_\mu\At^\mu_{(1)}=0,\labe{YMtlin-}\\
&&\square \At^i_{(1)}-\del^i\del_\mu\At^\mu_{(1)}=0,\labe{YMtlint}
\eeq
which are solved in Fourier space by 
\beq
&&-q^2\At^i_{(1)}(q)=-2 \left(\delta^{ij}-\frac{q^iq^j}{2q^+q^-}\right)\alphat^j_{_0}(\q)-2i\frac{q^i}{2q^+q^-}\betat_{_0}(\q),\nn
&&-q^2\At^-_{(1)}(q)=-\frac{2i}{q^+}\betat_{_0}(\q).\labe{res2}
\eeq

\section{Gluon production}
The spectrum of produced gluon is defined as follows 
\be
(2\pi)^32E\frac{dN}{d^3{\bf q}}=\sum\limits_\lambda \langle |{\mathcal M}_\lambda|^2\rangle,
\ee
where  $\lambda$ is the gluon polarization. The symbol $\langle...\rangle$ stands for the average over the color sources $\rho_{_A}$ and  $\rho_{_B}$\cite{GLV}. The amplitude ${\mathcal M}_\lambda$ is related to the classical gauge field by the reduction formula
\be
{\mathcal M}_\lambda=\lim_{q^2\rightarrow 0} q^2\At^i(q)\epsilon^i_\lambda(q) ,\labe{amp}
\ee
where $ \epsilon^i_\lambda(q)$ is the  polarization vector of the gluon and $q$ its four-momentum. In the axial gauge $A^+=0$, only the transverse components of the field contribute, and the sum over polarizations states is done with the help of the relation $\sum\limits_\lambda \epsilon^i_\lambda(q)\epsilon^{\ast j}_\lambda(q)=\delta^{ij}$.
Note that, for on-shell gluons ($2q^+q^--\q^2=0$), the condition of transversality $q_\mu M^\mu(q)=0$ is fulfilled, where $ M^\mu\equiv q^2\At^\mu(q)$.
By inserting  in  (\ref{amp}) the explicit expression of $\At^i_{(1)}(q)$ given in Eq.~(\ref{res2})  we get
\be
{\mathcal M}_\lambda=-2 \left(\epsilon_\lambda^j-\frac{\epsilon_\lambda\cdot\q}{\q^2}q^j\right)\alphat^j_{_0}(\q)-2i\frac{\epsilon_\lambda\cdot\q}{\q^2}\betat_{_0}(\q).\labe{res3}
\ee
 This   allows us to write the gluon spectrum in the following compact form \cite{BlaizM}:
\be
4\pi^3E\frac{dN}{d^3{\bf q}}=\frac{1}{\;\q^2}\langle |\q\times \balphat_{_0}(\q)|^2 +|\betat_{_0}(\q)|^2\rangle.\labe{resN}
\ee
Because of the color structure of the fields, it is not possible to write Eq. (\ref{resN}) in a $k_t$-factorized form in the case where the two nuclei are in the saturation regime. In case where one of the nuclei is dilute, one can expand Eq. (\ref{resN}) at first order in the weak source (this would be the case for a proton) we recover the well know $k_t$-factorization formula for proton-nucleus collisions \cite{BlaizGV1}.

\end{document}